\newcommand{\be}{\begin{eqnarray}}
\newcommand{\ee}{\end{eqnarray}}

\documentclass[a4paper]{jpconf}
\usepackage{graphicx}
\begin{document}
\title{Primordial Black Holes Around Us Now, 
Long Before, and Far away
%Preparing a paper using \LaTeXe\ for publication in  \jpcs
}

\author{A.D. Dolgov}

\address{Novosibirsk State University, Novosibirsk, Russia  
%Production Editor, \jpcs, \iopp, Dirac House, Temple Back, Bristol BS1~6BE, UK
}

\ead{dolgov@fe.infn.it}
%jacky.mucklow@iop.org}

\begin{abstract}
Recent astronomical data on Black hole observations are reviewed. The arguments in favor that the observed black holes are predominantly primordial (PBH) are presented. The mass spectrum of PBH is best fit to the log-normal one. A model of PBH formation with log-normal 
spectrum is briefly described.
\end{abstract}

\section{Introduction}

Astronomical observations of  the last decade present very strong evidence that the universe is filled with unexpectedly high amount of 
different kinds of black holes. Moreover these black holes are seen not only in the contemporary, almost 15 billion year old universe, but also
in the quite young one which was only 500 million year old.

The black holes are observed in all imaginable mass ranges:
{supermassive black holes (SMBH), $M=(10^{10} - 10^{6} ) M_\odot$,}
intermediate mass black holes (IMBH), ${M= (10^{2} - 10^5 ) M_\odot}$,
{black holes with masses of tens solar masses ${ M\sim 10 M_\odot}$,}
{and maybe black holes (BHs) even with a fraction of the solar mass.} 

These BHs live just next door in our Galaxy and in not so distant galaxies, as well as rather far-away but still in the present day universe,
and, what was a striking surprise, they were already formed during the first hundred million years after big bang, which is by far too short 
time according to the conventional scenario of their creation.

{The observed black holes may make all or a weighty fraction of the cosmological dark matter, 
have seed galaxy formation, and create binaries emitting gravitational waves  
observed at LIGO/Virgo interferometers.}

Most probably all, or almost all, those BHs are primordial (PBH). In this case the tension with the conventional cosmology and astrophysics,
created by their high abundance, smoothly disappears. 
These problems are reviewed in two-year old paper~\cite{AD-UFN}, but 
 a lot of new data since that time are accumulated indicating the same direction towards PBH..

{Recently a torrent  of {new abundant BHs, has been observed presumably primordial.}}
In any single case an alternative interpretation might be possible but the overall  picture
is very much in favor of massive {\it primordial black holes}

%Review: AD, Phys. Usp. 61 (2018) 2, 115-132; 

\section{Black holes by the formation mechanisms  \label{s-bh-form}}

{Three types of BH: astrophysical, accreting, primordial}\\[1mm]
{\it  I. Astrophysical BHs:} created by stellar collapse after star exhausted its nuclear fuel. Expected masses 
are just above the neutron star masses, ${\sim 3 M_\odot}$, and naively expected they should quite close to i
this value. Instead, the mass spectrum of BH in the Galaxy has maximum at 
${M \approx 8 M_\odot} $  with the width: $ \sim(1-2) M_\odot $. For the discussion of this problem and the 
list of references see e.g.~\cite{AD-UFN}. Bearing in mind that there are strong arguments that
the sources of the LIGO/Virgo observed gravitational waves are binaries of RBHs (see below sec. 4 ),
it is natural to accept that the black holes observed in the Galaxy are (mostly) primordial as well. \\[1mm]
{ \it II. BH created by matter accretion to excessive density regions. } \\
{There is a supermassive BH (SMBH) in any large galaxy with $M\geq 10^9 M_\odot$ in elliptic and lenticular 
galaxies and ${M\sim (10^6-10^7) M_\odot} $ in elliptic galaxies, as Milky Way.}
However, the known mechanisms of accretion are not efficient enough to create such monsters during the universe 
age ${t_U \approx 15 }$ Gyr.
{Very massive seeds are necessary, but their origin is mysterious.}
{Moreover SMBH are found in very small galaxies and one SMBH lives even in almost empty space.}

{SMBH are also observed recently with surprisingly large amount in quite young universe with the age about (1 - 0.5) Gyr.}
{Probably SMBH are primordial or created by the accretion to supermassive primordial seeds, see the next point.}\\[1mm]
{\it III. Primordial black holes (PBH)} created in the very early universe during pre-stellar epoch.\\
 The idea of the primordial black hole (PBH)  i.e. of black holes which could be 
formed  the early universe prior to star formation was pioneered by Zeldovich and Novikov~\cite{ZN-BH}
 %"The Hypothesis of Cores Retarded During Expansion and the Hot Cosmological Model", 
%Astronomicheskij Zhurnal, 43 (1966) 758,
 %Soviet Astronomy, AJ.10(4):602–603;(1967). \\
{ According to their idea, the 
density contrast in the early universe inside the piece of volume with the radius equal to the cosmological horizon 
might accidentally happen to be large, {${\delta \rho /\rho \approx 1}$,} then
that piece of volume would be inside its gravitational radius i.e. it became  a PBH, which
decoupled  from the cosmological expansion. }

The mechanism was elaborated and developed 
later by Hawking~\cite{SH-BH} and by Carr and Hawking~\cite{CH-BH}.

\section{Modified  mechanism of PBH formation}

An essentially different mechanism for creation of primordial black holes was suggested in our work of 
1993~\cite{AD-JS}, see also an extension and more detail in ref.~\cite{DKK}. The mechanism is based on the popular 
Affleck-Dine (AD) scenario of baryogenesis~\cite{AD-BG} realized at inflationary epoch and immediately after. The AD
baryogenesis is stimulated by supersymmetry (SUSY) at high energy scale. In such a theory must exist a scalar field with non-zero
baryonic number, we call it $\chi$. The potential of $\chi$ possesses the so called flat directions, along which the potential
does not rise but remains constant. Due to quantum fluctuations during inflation field $\chi$ may "travel" far along one or other
flat direction and acquire a large amplitude.

After inflation was over and SUSY broke, so that $\chi$-field acquired non-zero mass, the flat directions became curved,  and $\chi$
started to roll down towards the origin, $\chi =0$. On the way down field $\chi$ gained huge baryonic number, which finally led to even 
large baryon asymmetry of the universe which may be even of order unity, much larger than the observed  one
$\beta = 6\cdot 10^{-10}$. So theorists have to  invent some clever ways to suppress the asymmetry down to the proper value.

The new PBH creation mechanism could be realized if  ${\beta}$ reached large values 
only in cosmologically small but possibly astronomically large bubbles, 
while in the bulk of the universe it has normal value. 
{${{ \beta \approx 6\cdot 10^{-10}}}$.}  This may be achieved by introduction of the
general renormalizable coupling of the AD baryonic scalar field with inflaton, see below eq.~(\ref{U-Phi-chi}).

{The fundament of PBH creation is set on at inflation by making large isocurvature
fluctuations at relatively small scales, with practically vanishing density perturbations.} %\\[1mm]
The huge perturbations in baryonic number transformed later into density perturbations at the
QCD phase transition when massless quarks turned into heavy baryons.

{The emerging universe looks like a piece of Swiss cheese, where holes are high baryonic 
density bubbles (HBB) occupying a minor fraction of the universe volume.} 
Inflationary prehistory allows for  creation of  huge PBH with masses up to  ${(10^4-10^5) M_\odot}$,
or even higher depending on the model. The  mass spectrum of the created PBHs has very simple
log-normal form with only 3 constant parameters: ${\mu}$, ${\gamma}$,  ${M_0}$:
\be
\frac{dN}{dM} = \mu^2 \exp{[-\gamma \ln^2 (M/M_0)].}
\label{dn-dM}
\ee
The values of ${\gamma}$ and ${\mu}$ depend upon unknown high energy physics at the AD baryogenesis,
{but the central mass, ${M_0}$, is equal to the known mass inside horizon at the QCD phase 
transition~\cite{AD-KP-mass}.

 The mass inside horizon at RD stage, ${r_{hor} = 2 t}$ is
 \be 
 M_{hor} = m_{Pl}^2 t  .
 \label{M-hor-RD}
 \ee	
If ${ \delta \rho /\rho = \kappa}$, then ${M_{BH} = \kappa M_{hor}}$ and the gravitational radius is
\be 
r_g = \frac{2M}{m_{Pl}^2} = 2 \kappa r_{hor}.
 \label{r-g}
 \ee
 For PBHs formed at the QCD phase transition at $T\sim 100 $ MeV, and the universe age equal to
  $t = 4\cdot 10^{-5} \left({100\,{\rm MeV}}/{T}\right)^2\,{\rm sec} $ we find
 \be
 M_{hor}  = 8 M_\odot \cdot  \left(\frac{100\,{\rm MeV}}{T}\right)^2. 
 \label{M-hor}
\ee
According to lattice calculations $T_{QCD} = 100 -150 $ MeV but if quark chemical potential is large,
$T_{QCD} $ may be smaller and ${M_0}$ be bigger.

So the central mass of PBH log-normal mass spectrum is predicted to be close to
 ${10 M_\odot}$~\cite{AD-KP-mass}  
 %(AD, K.Postnov, JCAP 07 (2020) 063, astro-ph 2004.11669 ) }
  As we see in what follows, this result is  
 in good agreement with observations, see figures 1 and 2 below.

{Such form of the mass spectrum and similar ones, the so called extended spectra,  became quite popular nowadays.}
The suggested scenario of PBH formation~\cite{AD-JS} pioneered in implementation of
{inflation} to PBH formation. It allows for { PBH huge masses, much larger than horizon 
mass in the very early universe.} 
To the present time a  long list of works on inflationary formation of PBH  came to life.

Let us describe a toy model which has the desired properties. The quartic potential of $\chi$ with flat directions  
can be presented as:
\be
U_\lambda(\chi) = \lambda |\chi|^4 \left( 1- \cos 4\theta \right)
\ee
and of the mass term,  i.e. quadratic potential may have the form:
\be
U_m( \chi ) = m^2 |\chi|^2} {\left[{ 1-\cos (2\theta+2\alpha)}  \right],
\ee
where ${\chi = |\chi| \exp (i\theta)}$ and ${m=|m|e^\alpha}$.
{If ${\alpha \neq 0}$, C and CP symmetries would be explicitly broken.}

In grand unified (GUT) SUSY models baryonic number is naturally non-conserved. In our toy model the non-conservation of baryons 
is enforced  by non-invariance of potential ${U(\chi)}$ with respect to the phase rotation, $\chi \rightarrow \chi \exp ( i \phi )$. Deviation
form thermal equilibrium is practically evident. 
So all three Sakharov's conditions for baryogenesis are fulfilled.

{{Initially (after inflation) ${\chi}$ could be away from origin and, when 
inflation is over, started to evolve down to equilibrium point, $\chi = 0$
according to the equation similar to that of  theNewtonian mechanics:}}
\be
\ddot \chi +3H\dot \chi +U' (\chi) = 0.
\ee
Baryonic number of $ \chi$:
\be
B_\chi =\dot\theta |\chi|^2
\label{B-chi}
\ee
is analogous to mechanical angular momentum. 
${{\chi}}$ decays transferred
baryonic charge to that of quarks in B-conserving process.

If $ { m\neq 0}$, 
the angular momentum, or B (\ref{B-chi}), is generated by a different 
direction of the  quartic and quadratic valleys at low ${\chi}$.
{If CP-odd phase ${\alpha}$ is small but non-vanishing, both baryonic and 
antibaryonic domains might be  formed}
{with possible dominance of one of them.}\\
{Matter and antimatter domains may exist but globally cosmological excess of baryons over antibaryons (or vise versa)  is more probable 
than globally B-symmetric universe.
 
Our new input to the model is  an introduction of the Affleck-Dine field $ \chi$  couplng to inflaton $ \Phi$, the first term in the 
equation below:
\be 
U = {g|\chi|^2 (\Phi -\Phi_1)^2}  +
\lambda |\chi|^4 \,\ln \left( \frac{|\chi|^2 }{\sigma^2 } \right)
+\lambda_1 (\chi^4 + h.c. ) + 
(m^2 \chi^2 + h.c.). \,\,\,\,\,\,\,\,\,\,\,\,\,\,\,\,\,\,\,\,
\ee
As we have already mentioned, C and  CP would be broken, if the relative phase of ${\lambda_1}$ and 
$m$ is non-zero, otherwise one can  "phase rotate'' $ \chi$ and come to real coefficients in the potential above. 
%{Coupling to fermions may break CP.}\\

The introduced above interaction between $\Phi$ and $\chi$ is quite natural.
An interaction between two scalar fields  ${\Phi}$ and ${\chi}$ in one or other form  must exist.
This coupling is a general renormalizable one. 
{The only mild tuning is that ${\Phi}$ reached and passed ${\Phi_1}$  during inflation.}
Duration of inflation after that is a free parameter.

When $\Phi$ is close to $ \Phi_1 $, the window to flat direction is open but presumably 
only during  a short period, cosmologically small but possibly astronomically large 
bubbles with high ${ \beta}$ could be
created, occupying {a small
fraction of the universe,} while the rest of the universe has normal
{${{ \beta \approx 6\cdot 10^{-10}}}$, created 
by small ${\chi}$}. The formed in this way HBBs would turn in massive black holes after the QCD phase transition.
However, HBBs with considerably smaller masses may not forrn black holes but some dense stellar-like objects.
%{Phase transition of 3/2 order.}\\

{This mechanism of massive PBH formation is quite different from all others.}
{The fundament of PBH creation is build at inflation by making large isocurvature
fluctuations at relatively small scales, with practically vanishing density perturbations.}
The initial isocurvature perturbations are created by the
 large variation of chemical potential of massless quarks and antiquarks.
Density perturbations are generated rather late after the QCD phase transition
when quarks turn into massive baryons.
The emerging universe may be full of massive and supermassive black holes.

{The outcome of the discussed mechanism, depending on ${\beta = n_B/n_\gamma}$ could be:}
\begin{itemize}
\item
PBHs with log-normal mass spectrum.
\item
{Compact stellar-like objects, similar e.g. to cores of red giants.}
\item
{Disperse hydrogen and helium clouds  with (much) higher than average ${n_B}$ density.}
\item
{${\beta}$ may be negative leading to compact antistars which could survive annihilation with the 
homogeneous baryonic background.}
\end{itemize}
A modification of inflaton interaction with scalar baryons as e.g.
\be 
U \sim |\chi|^2 (\Phi - \Phi_1)^2 ((\Phi - \Phi_2)^2
\label{U-Phi-chi}
\ee
gives rise to a superposition of two log-normal spectra or multi-log. The consequences of this modification are not yet explored.

\section{Gravitational waves and PBHs \label{s-GE-PBH} }
 
 %Gravitational waves from BH binaries. Chirp mass 

{Two rotating gravitationally bound massive bodies are known to emit gravitational 
waves.} In quasi-stationary inspiral regime, the radius of the orbit and the rotation frequency
slowly change and the GW frequency is approximately twice the Newtonian rotation frequency:
\be 
\omega^2_{orb} =  \frac{M_1+M_2}{m_{Pl}^2 R^3}\,.
\label{omega-GE}
\ee

{The luminosity of the GW radiation during this stage is:}
\be 
L  = \frac{32}{5}\,m_{Pl}^2\left(\frac{M_c\,\omega_{orb}}{m_{Pl}^2}\right)^{10/3}\,,
\label{GW-lum}
\ee
where $M_1$, $M_2$ are the masses of two bodies in the binary system and 
${M_c}$ is the so called chirp mass: 
\be 
M_c=\frac{(M_1\,M_2)^{3/5}}{(M_1+M_2)^{1/5}} \, ,
\label{M-chirp}
\ee

Discovery of gravitational waves  (GW) by LIGO strongly indicate that the sources of GW are primordial black holes,
 see e.g.~\cite{BDPP}. The conventional astrophysical black holes are much less favorable by the following reasons:\\
{1. An astrophysical formation of very massive BHs,  (${M \sim 30 M_\odot}$),} demands much more massive progenitors 
with $M > 100 M_\odot$. but they are not observed in sufficient  amount. Recently the problem of stellar origin of the
LIGO sources became multifold more pronounced after observation of the event GW190521, see fig. 3. On other hand, 
possible existence of so heavy progenitors can be explained by an extra energy source due to annihilation of DM indise
stars~\cite{KF-ann}.\\
{2. Difficulties with formation of BH binaries from the original stellar binaries. 
{If BH is created through stellar collapse,} {a small non-sphericity results in a huge recoil momentum
of the BH and the binary would be destroyed.} 
{The problem of the binary formation is simply solved if the observed sources of GWs are the binaries of
primordial black holes.} 
{They were at rest in the comoving volume, when inside horizon they are gravitationally attracted and  
may loose energy due to dynamical friction in the early universe.
The probability %of capture is have non-negligible probability
to become gravitationally bound for PBHs is non-negligible.}
The conventional scenario is not completely excluded but seems much less probable.\\
3.  The low value of the angular momenta of the original BHs in the coalescing binaries is difficult to 
understand if they are the usual astrophysical BHs formed in the process of stellar collapse.
{Still, individual non-rotating PBHs forming a binary initially rotating on elliptic orbit could gain collinear spins about 0.1 - 0.3,
rising with the PBH masses and eccentricity~\cite{PM-spin,PKM-spin} 
{This result is in agreement with the most massive events 
GW170729 LIGO  produced by the binary with masses ${50 M_\odot}$  and ${30 M_\odot}$ and
and  GW190521 produced by BHs with ${85 M_\odot}$  and ${65 M_\odot}$}.

{ To summarize: each of the mentioned problems might be solved in the conventional frameworks but it looks
much simpler to assume that the LIGO sources are primordial.}

A very strong argument in favor of PBH sources of the registered gravitational waves  is presented  in 
ref.~\cite{DKMP} on the basis of the analysis of the chirp mass distribution found from all publicly available LIGO events,
namely, the available data on the chirp mass distribution of the black holes in the coalescing binaries in O1-O3 LIGO/Virgo runs are
analyzed and compared with theoretical expectations based on the hypothesis that these black holes are primordial with 
log-normal mass spectrum. The inferred best-fit mass spectrum parameters, $M_0=17 M_\odot$ and $\gamma=0.9$, fall 
within the theoretically expected range and shows excellent agreement with observations as is presented in figs. 1 and 2.
In fig. 3 the data on the masses of the LIGO observed binaries, both the masses of the initial members of the binaries and  on the 
mass of the resulting BH are presented. With the last event, GW190521,  
not included into figs. 1 and 2, the agreement becomes noticeably better. 

On the opposite, binary black hole models based on massive binary star evolution require serious 
additional adjustments to reproduce the observed chirp mass distribution, see fig. 4.

\begin{figure}[htbp]
\begin{center}
\includegraphics[scale=0.2]{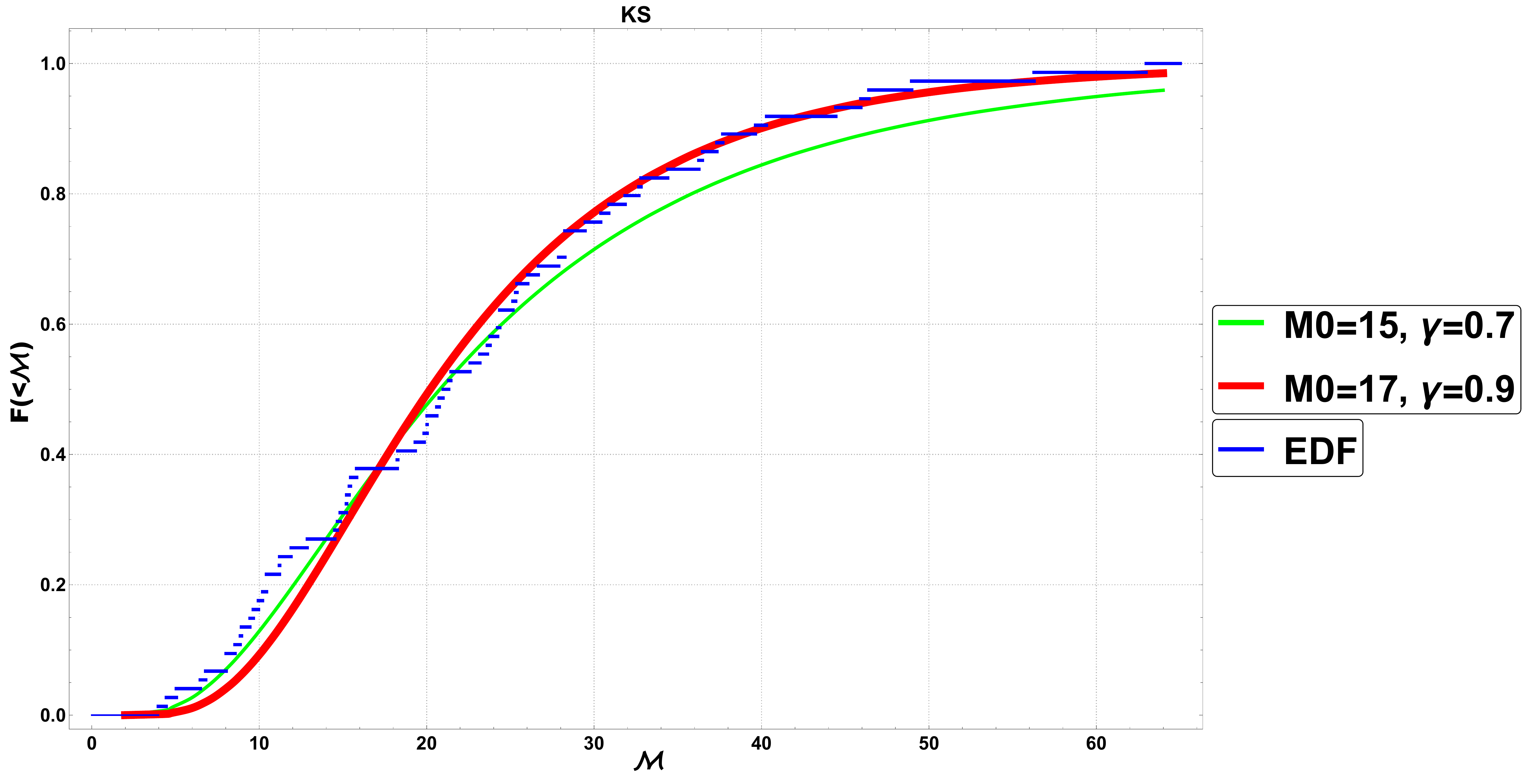}
\caption{Model distribution $F_{PBH}(< M)$ with parameters  $M_0$ and $\gamma$ for two best 
Kolmogorov-Smirnov tests.  EDF= empirical distribution function.}
\end{center}
\end{figure}

\begin{figure}[htbp]
\begin{center}
\includegraphics[scale=0.2]{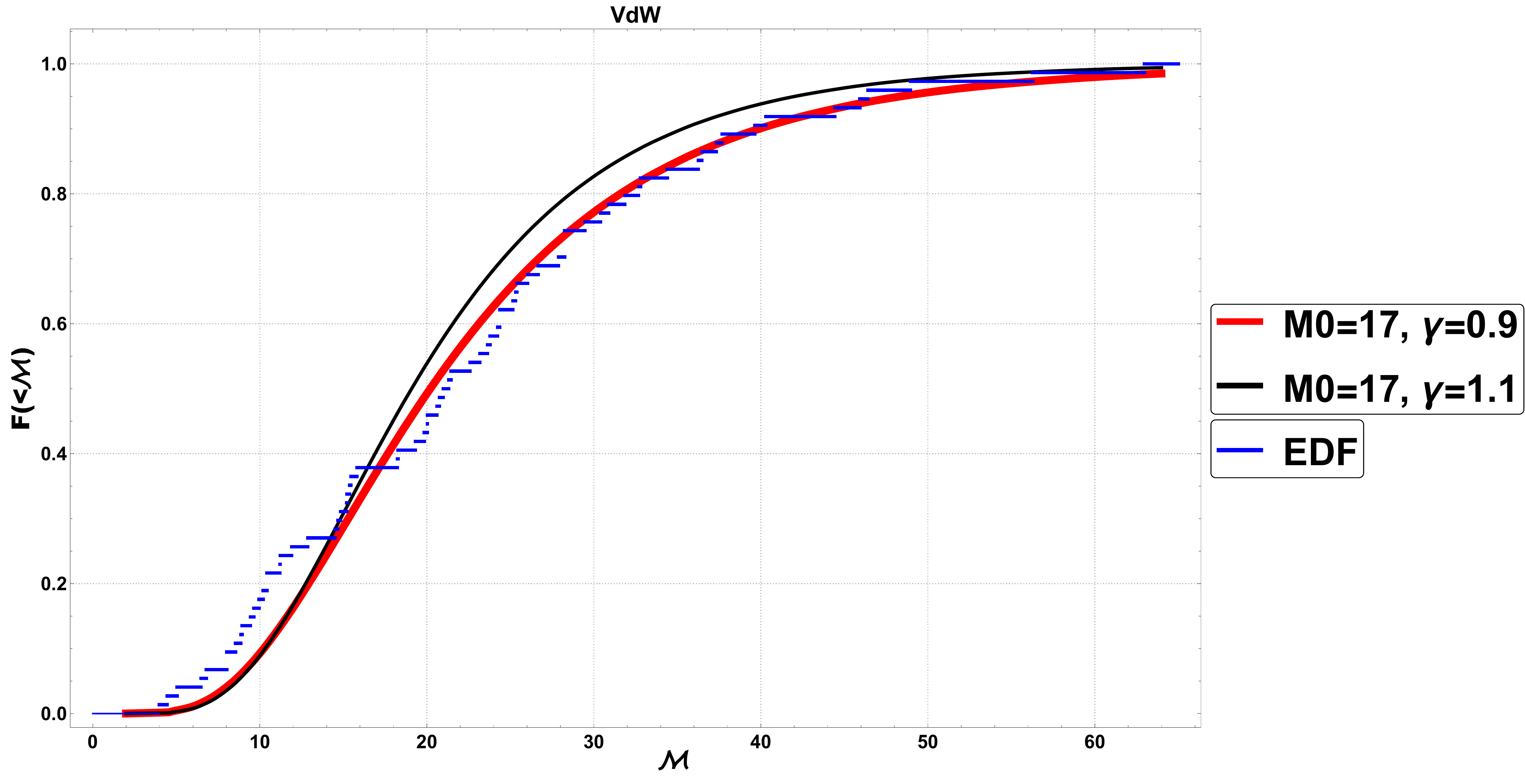}
\caption{Model distribution $F_{PBH}(< M)$ with parameters  $M_0$ and $\gamma$
for two best Van der Waerden tests.}
\end{center}
\end{figure}

%{ Merging of ${(65+85) M_\odot \rightarrow 142 M_\odot }$.}\\
%{ With this event the agreement is even better. }
%The spins of the initial BHs may be about 0.3 as predicted by Postnov and Mitichkin.
\begin{figure}[htbp]
\begin{center}
\includegraphics[scale=0.5]{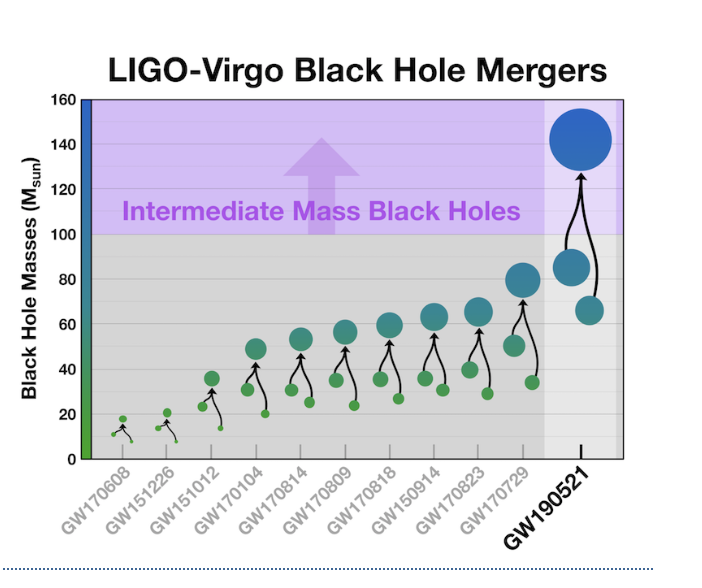}
\caption{ Merging of ${(65+85) M_\odot \rightarrow 142 M_\odot }$.}
\end{center}
\end{figure}

\begin{figure}[htbp]
\begin{center}
\includegraphics[scale=0.2]{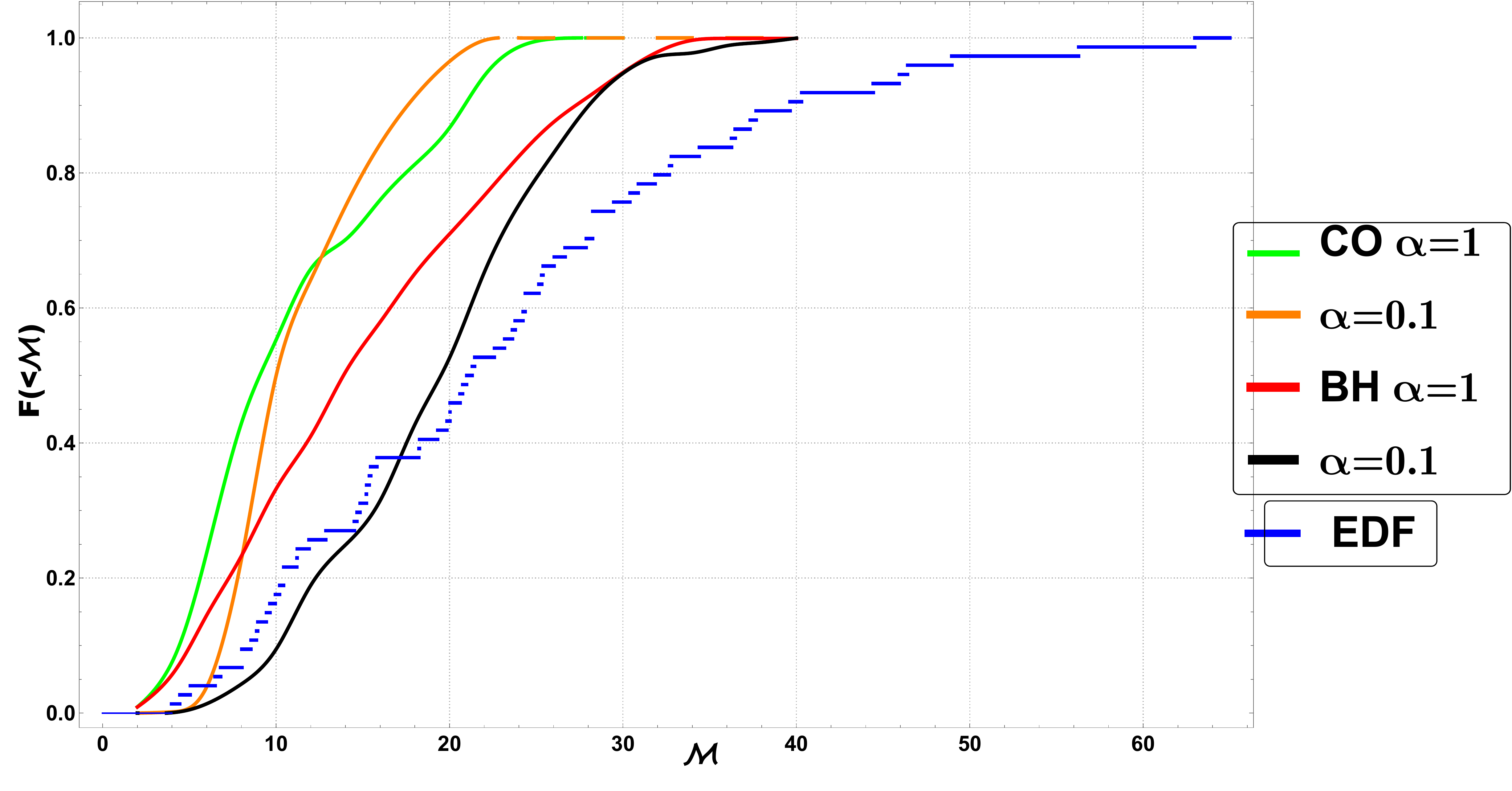}
\caption{Cumulative distributions $F(< M)$ for several { astrophysical} models of binary BH coalescences. }
\end{center}
\end{figure}
 
% \end{document}
 
\section{SMBH in contemporary universe \label{s-smbh-today}}

{Every large galaxy available to the proper study 
contains a central SMBH with
mass {${\sim(10^6-10^7) M_\odot}$} in spiral galaxies like Milky Way
 and larger than}  { ${ 10^{9}M_\odot}$} in giant elliptical
and compact lenticular galaxies, up to the record {{66 billions solar masses, TON 618~\cite{ton-618}}. 
{The origin of these BHs is mysterious. 
The accepted faith is that
these BHs are created by matter accretion to a central seed. 
{But, the usual accretion efficiency is insufficient to create them during the Universe life-time,
14.6 Gyr.} 

According to ref.~\cite{murch} building up SMBH SgrA* with the mass ${\sim 4\times10^6 M_\odot}$ residing at the 
centre of our galaxy. within  14.6 billion %${\sim 10^{10}}$ 
year lifetime of our galaxy would require a mean accretion rate of  ${4\times 10^{-4}} M_\odot$  per year.
At present, X-ray observations constrain the rate of hot gas accretion 
to ${\dot{M} \sim 3 \times 10^{-6} M_\odot}$ per year 
and polarization measurements constrain it near the event horizon to 
${\dot{M}_{hor} \sim 10^{-8} M_\odot}$/yr. 
{The universe age is short by two orders of magnitude.}

{Even more puzzling is that SMHBs  are observed  in  { very small galaxies}
 and even in almost empty space, where no material to make a SMBH can be found~\cite{naked-BH}.

{The mass of BH is typically 0.1\% of the mass of the stellar bulge of galaxy} 
% \cite{BH-bulge,*SaniMarconi}
but some galaxies may  have huge BH: {e.g. NGC 1277  has
the central BH  of  ${1.7 \times 10^{10} M_\odot}$, or ${60}$\% of its bulge mass.}
%\cite{NGC1277}.
This creates serious problems for the
scenario of formation of central supermassive BHs by accretion of matter in the central 
part of a galaxy~\cite{khan}.

A few more examples of SMBH which are at least an order of magnitude more massive than their  
host galaxy suggests are Henize 2-10, NGC 4889,
and NGC1277. 

An inverted picture seems top be much more plausible, when first a supermassive BH was formed and 
attracted matter seeding the galaxy formation, as advocated in refs~\cite{AD-JS,DKK,Bosch}.

Another possible piece of evidence in favor of primordial formation of SMBHs is their clumping. There are
at least four binaries of SMBH, one triple system and one quartet. A list of references can be found in  the
review~\cite{AD-UFN}. 

An orthodox point of view which might explain the SMBH binary existence is 
{merging of two spiral galaxies creating an elliptical
galaxy, leaving two or more SMBHs in the center of the merged elliptical.}
The traditional way of  formation of a triple systems of SMBHs to say nothing of the quartet is much more difficult.
Heretic but simpler possibility is that primordial SMBHs formed binaries, triple systems, and quartet  
in the very early universe.

\section{SMBH in young universe \label{smbh-early}}

In the early universe at high redshifts ${z> 6}$
about 100 QSO are known, corresponding to billion solar mass  SMBH. The QSO with maximum redshift: z=7.54 with 
800 million solar masses is observed by the group~\cite{max-z}. Note that this QSO is situated in the neutral universe, in other words
the interstellar medium was not ionized. This means that the accretion was absent or insignificant.

Another very interesting observation is that of the second largest QSO which is powered by 
1.5 billion of solar masses~\cite{2nd-max}.
According to the author's statement,  
models indicate it must have {formed not later than 100 million years after the Big Bang, extremely short time for its creation.

In addition to that another monster of 12 billion solar mass was discovered~\cite{max-mass}
The problem with formation of lighter quasars  multifold deepens with this new fantastically massive "creature".
%\tcred{The new one with $\bm{M \approx 10^{10} M_\odot }$  makes the formation}
%\tcblue{absolutely impossible in the standard approach.} %\\[3mm]
On the other hand the accretion rate according to the conventional estimates~\cite{accr-rate} 
is extremely low,
%M.A. Latif, M Volonteri, J.H. Wise, [1801.07685] 
a halo with the  mass of ${ 3 \times 10^{10}~M_{\odot}}$ at ${z=7.5}$;
accretes only about 2200 ${M_{\odot}}$ during 320 Myr.

Recently a striking observation has been done~\cite{short-time} 
that the life-time of activity of QSO at $z\approx 6$ is only $10^3 - 10^4$ years.
It means that  only a minor fraction of QSO, and thus of SMBH, is observed, about $10^{-4}$ or even less.
So the early unverse is really overpopulated  by SMBH.

The universe at redshifts $z= 5-10$ is also unexpectedly rich of massive and luminous galaxies, with
the luminosity up to {${L= 3\cdot 10^{14} L_\odot }$. The corresponding  list of references can be found in~\cite{AD-UFN}
According to ref.~\cite{early-gal}
 the {density of galaxies at ${z \approx 11}$ is 
${10^{-6} }$ Mpc${^{-3}}$, an order of magnitude higher than estimated from the data at lower z.}
{The origin of these galaxies is unclear.}
So, again the conclusion is almost unavoidable that the inverted picture of galaxy formation can solve the problem: primordial SMBHs seeded galaxies but  not vice versa, and not only in young universe but also today.

}

To conclude on QSO/SMBH,
the quasars are supposed to be supermassive black holes
{and their formation in such short time by conventional mechanisms looks problematic.} 
{Such black holes,
{when the Universe was less than one billion years old,} 
present substantial challenges to theories of the formation and growth of
black holes and the coevolution of black holes and galaxies.}
{Even the formation of SMBH in contemporary universe during 14 Gyr  is hard to explain.} 
{Non-standard accretion physics and the formation of massive seeds seem to be necessary.
Neither of them is observed in the present day universe.}

{It is difficult to understand how {${10^9 M_\odot}$} black holes  {(to say nothing about ${10^{10} M_\odot}$)}
appeared so quickly after the big bang {without invoking non-standard accretion physics
and the formation of massive seeds,}} 
{both of which are not seen in the local Universe.}

There are more problems in contemporary universe and in the universe 14 billion years ago. Because of lack of space and
time they are simply enumerated here:\\
{$\bullet$ MACHOS: invisible stellar type objects with ${M \sim 0.5 M_\odot}$
%observed through gravitational microlensing.}
\\
{$\bullet$ IMBH unexpected but quite abundant, thousands of them.}\\
{$\bullet$ Peculiar stars" too old, too fast, with strange chemistry.}\\
{$\bullet$ Globular clusters seeding by ${ (10^3 - 10^4) M_\odot}$ BHs,
 and dwarfs by ${ (10^4 - 10^5) M_\odot}$ BHs~\cite{AD-KP-glob}.} 
% AD, K.Postnov, JCAP 04 (2017) 036. \\
{$\bullet$ Overpopulation of the young, ${ z\sim 10}$  universe with early created 
gamma-bursters and supernovae, early bright galaxies, evolved chemistry including huge amount of dust.}

%{Intermediate Summary}
%{Model predictions, postdictions and claims.}\\

\section{Conclusion \label{s-concl}}

Here we summarize the basic features of the model of refs.~\cite{AD-JS,DKK} , preiditons, postdictions, and 
development in subsequent papers.\\
 %\begin{itemize}
%\item
{$\bullet$ 1. Natural model based on AD-baryogenesis scenario
leads to abundant formation of PBHs and compact stellar-like
objects in the early universe after QCD phase transition, $ { t \geq 10^{-5}} $ sec.} \\
%\item
{ $\bullet$ 2. Thess compact objects have simple log-normal mass spectrum.} \\
%\item
{$\bullet$  3. PBHs formed at this scenario can explain the peculiar features of the sources
of GWs observed by LIGO.}}\\
%\item
{$\bullet$
 4. The considered mechanism solves the numerous mysteries of ${z \sim 10}$ universe: abundant population
of supermassive black holes, 
early created 
gamma-bursters and supernovae, early bright galaxies,and evolved chemistry including dust}.} \\
%\item
{$\bullet$ 5. There is persuasive data in favor of the inverted picture of galaxy formation, when first a supermassive 
BH seeds  are formed and later they accrete  matter forming galaxies.}\\
%\end{itemize}
{$\bullet$ 6. An existence of supermassive black holes observed  in all large and some small galaxies 
and even in almost empty environment is naturally explained.}\\
{$\bullet$ 7. "Older than ${t_U}$" stars may exist; the 
large age is mimicked by the unusual initial chemistry. }\\
%\item
%\tcblue{$\bullet$ 8. Existence of high density invisible "stars" (machos) is understood.} \\
%\item
{ $\bullet$  8. Explanation of the origin of BHs with 2000 ${M_\odot}$ in the core of globular cluster (GC) and the
observed density of GC is presented.}}\\
%\item
$\bullet$ 9. A large number of the recently observed IMBH was  predicted.\\
%\item
{$\bullet$ 10. A large fraction of dark matter or even 100\% can be made of massive PBHs.}\\
%\item
 {$\bullet$ 11. Clouds of matter with high baryon-to-photon ratio and unusual chemkistry can exist.}}\\
%\item
{$\bullet$ 12. A possible by-product: plenty of (compact) anti-stars, even in the Galaxy,}
{not yet excluded by observations}.  %(C. Bambi, S. Blinnikov, AD., K.Postnov).\\
%\end{itemize}
Extreme point of view:\\
{$\bullet$ {Black holes in the universe are mostly primordial (PBH).}}\\
$\bullet$ {Primordial BHs make all or dominant part of dark matter (DM).}

\section*{Acknowledgments}
The work was supported by the Ministry of Science and Higher Education grant\\  No.  FSUS-2020-0039

\end{document}